\documentclass[aps,twocolumn,groupedaddress,superscriptaddress,amsfonts,amssymb,amsmath,citeautoscript,a4paper]{revtex4}

\usepackage{siunitx}
\usepackage{graphicx}

\begin{document}

\title{The importance of substrates for the visibility of "dark" plasmonic modes}

\author{Saskia~Fiedler}
\email{safi@mci.sdu.dk}
\affiliation{Center for Nano Optics, University of Southern Denmark, Campusvej 55, DK-5230~Odense~M, Denmark}

\author{S{\o}ren~Raza}
\affiliation{Department of Physics, Technical University of Denmark, DK-2800 Kgs. Lyngby, Denmark}

\author{Ruoqi~Ai}
\affiliation{Department of Physics, The Chinese University of Hong Kong, Shatin, Hong Kong SAR, China}

\author{Jianfang~Wang}
\affiliation{Department of Physics, The Chinese University of Hong Kong, Shatin, Hong Kong SAR, China}

\author{Kurt~Busch}
\affiliation{Max-Born-Institut, Max-Born-Stra{\ss}e 2A, 12489 Berlin, Germany}
\affiliation{Humboldt-Universit{\"a}t zu Berlin, Institut f{\"u}r Physik,
D-12489 Berlin, Germany}

\author{Nicolas~Stenger}
\affiliation{Department of Photonics Engineering, Technical University of Denmark, DK-2800 Kgs. Lyngby, Denmark}
\affiliation{Center for Nanostructured Graphene, Technical University of Denmark, DK-2800 Kgs. Lyngby, Denmark}

\author{N.~Asger~Mortensen}
\affiliation{Center for Nano Optics, University of Southern Denmark, Campusvej 55, DK-5230~Odense~M, Denmark}
\affiliation{Danish Institute for Advanced Study, University of Southern Denmark, Campusvej 55, DK-5230~Odense~M, Denmark}
\affiliation{Center for Nanostructured Graphene, Technical University of Denmark, DK-2800 Kgs. Lyngby, Denmark}

\author{Christian~Wolff}
\affiliation{Center for Nano Optics, University of Southern Denmark, Campusvej 55, DK-5230~Odense~M, Denmark}
\email{cwo@mci.sdu.dk} 

\begin{abstract}
Dark plasmonic modes have interesting properties, such as a longer lifetime and a narrower linewidth than their radiative counterpart, as well as little to no radiative losses. However, they have not been extensively studied yet due to their optical inaccessibility.
Using electron-energy loss (EEL) and cathodoluminescence (CL) spectroscopy, the dark radial breathing modes (RBMs) in thin, monochrystalline gold nanodisks are systematically investigated in this work. It is found that the RBMs can be detected in a CL set-up despite only collecting the far-field. Their visibility in CL is attributed to the breaking of the mirror symmetry by the high-index substrate, creating an effective dipole moment. The outcoupling into the far-field is demonstrated to be enhanced by a factor of \num{4} by increasing the thickness of the supporting SiN membrane from \SI{5}{\nano\meter} to \SI{50}{\nano\meter} due to the increased net electric dipole moment in the substrate.
Furthermore, it is shown that the resonance energy of RBMs can be easily tuned by varying the diameter of the nanodisk, making them promising candidates for nanophotonic applications.
\end{abstract}

\maketitle

\section{Introduction}

The optical properties of metallic nanoparticles (NPs) are well-known to be dominated by localized surface plasmons (LSPs) -- collective electron oscillations which produce an evanescent optical field, confined at the NP. As a result, the electric field at the surface of the plasmonic nanostructure is greatly enhanced. The optical absorption is highest at its resonance energy which can be tailored by varying the shape, the dimension, the dielectric surrounding, as well as by the choice of material such as highly doped semiconductors or metals ~\cite{Kelly:2003,Renwen:2017,Christensen:2017}. For metallic NPs smaller than the wavelength of the incident light, the strongest optical response is typically the resonance associated with the dipolar plasmonic mode at a frequency, $\omega$, below the plasma frequency, $\omega_p$. Nevertheless, for $\omega<\omega_p$, there can also exist higher order modes which can be either bright -- they possess an effective dipole moment for the electromagnetic field to couple to -- or dark -- they do not have a net dipole moment and thus cannot be excited by plane-wave light at normal incidence~\cite{Raza:2015, Chu:2009}. 
Flat plasmonic NPs with sufficiently high symmetry such as the the circular disks studied here can further support dark radial breathing modes (RBMs).
These RBMs should not be confused with radial bulk plasma oscillations e.g. in spheres~\cite{Raza:2011,Christensen:2014,Raza:2015b,Hille:2016}, which have a deceptively similar radial profile, but are due to the nonlocal response of metals above the plasma frequency ($\omega>\omega_p$).
In contrast, the RBMs studied in the following are surface plasmon polaritons at frequencies below the plasma frequency ($\omega<\omega_p$), only associated with a radial variation of the surface polarization charge and can be adequately described by a local material response such as a Drude--Lorentz model.
RBMs do not posses a net dipole moment which is why they are optically inaccessible with light at normal incidence. However, they can be excited by an electron beam in a scanning transmission-electron microscope (STEM), taking advantage of the near-field excitation and detection using electron-energy loss (EEL) spectroscopy~\cite{Chu:2009, Schmidt:2018, Schmidt:2012}. 
In conjunction with the electron-energy loss spectrum, this technique allows to probe the electromagnetic local density-of-states (LDOS) of plasmonic nanostructures projected to the swift electron's trajectory~\cite{GarciadeAbajo:2010, Kociak:2014,Polman:2019,GarciadeAbajo:2008}.
In the case of thin structures --- such as the one studied in this manuscript --- this provides a very good idea of the spatial distribution of the LDOS experienced by a vertically aligned emitter and therefore of the plasmonic mode structure. 
EEL spectroscopy is a very powerful tool to study plasmonic nanostructures with a sub-nanometer spatial and high spectral resolution~\cite{Cube:2013,Walther:2016}, allowing to correlate structural with spectral features at the nanoscale regardless of their "brightness". However, it has recently been shown that dark modes in plasmonic NPs can also be made visible by taking advantage of retardation effects in sufficiently large nanostructures, using cathodoluminescence (CL) or optical spectroscopy at an oblique angle~\cite{Schmidt:2012, Schmidt:2018, Krug:2014}.
Another way to get more insight in dark plasmonic modes has been achieved by breaking the symmetry, either by asymmetrically arranging metallic nanostructures or breaking the symmetry of an individual NP~\cite{Hao:2008, Panaro:2014}. Nevertheless, dark modes have not been extensively studied, or fully understood, yet despite their interesting optical properties such as a narrower line width, and a longer lifetime than bright modes which make them promising candidates for future nanophotonics~\cite{Fernandez-Dominguez:2017}, including sensing~\cite{Brolo:2012}, structural colors~\cite{Kristensen:2017}, metasurfaces~\cite{Ding:2018}, and enhanced light-matter interactions for quantum plasmonics~\cite{Fernandez-Dominguez:2018}. 

In this work, we investigate RBMs in individual monocrystalline Au nanodisks (NDs) utilizing EEL and CL spectroscopy, as well as numerical calculations of their electrodynamic response using the discontinuous Galerkin time-domain method (DGTD)~\cite{Busch:2011,Matyssek:2011} to gain more insight in the coupling of the dark modes to the far-field. These techniques are used in a complementary manner: EEL spectroscopy probes the near-field of the Au ND, allowing to identify the spatial and spectral properties of the dark RBM. However, it does not give any information on how or to what extent these dark modes can couple to the far-field. On the other hand, CL spectroscopy combines the near-field excitation by an electron beam with a far-field collection of their radiation component. To experimentally study the coupling of the RBMs to the far-field, CL is, therefore, preferable over optical spectroscopy which can only excite and detect in the far-field, typically resulting in a extremely low  or no signal of dark plasmonic modes even at an ideal excitation and detection angle.

Our experimental data show that dark plasmonic modes, namely RBMs, in Au NDs can not only be detected by EEL but also by CL spectroscopy, despite probing only the radiative part of the LDOS. In contrary to previously reported CL measurements of RBMs in silver NDs~\cite{Schmidt:2018}, this work reveals CL signal from RBMs in Au NDs with much smaller diameters down to \SI{110}{\nano\meter}. This suggests that retardation effects, as proposed by Schmidt \emph{et al.}~\cite{Schmidt:2018}, cannot fully explain our results. Therefore, we present a systematic study of RBMs in Au NDs and put forward a different explanation for their visibility in CL, involving symmetry breaking by the presence of a thin high-index substrate. 

\section{Methods}

\subsection{Au nanodisk sample preparation}
Thin monocrystalline nanoplatelets are produced by a seed-mediated method which is followed by an oxidation process, etching primarily under-coordinated atoms at the corners. By carefully controlling the process parameters, the diameter of the resulting Au NDs can be varied over a large range, while their thickness is determined by that of the initial nanoplatelet. More details can be found elsewhere~\cite{Cui:2018}. In this work, the Au NDs' diameter ranges from \SI{110}{\nano\meter} to \SI{195}{\nano\meter} with a thickness of \SI{20}{\nano\meter}. 
The Au NDs, stored in a cetyltrimethylammonium bromide (CTAB) solution, are centrifuged and subsequently diluted in DI water twice before being drop-casted onto \SI{5}{\nano\meter} to \SI{50}{\nano\meter} thin SiN membranes (purchased from TEMwindows). 
It is noteworthy that the Au NDs are still capped with an approximately \SI{1}{\nano\meter} thin layer of CTAB. The Au NDs have high optical quality with relatively low intrinsic Ohmic losses enabled by their monocrystallinity, which was recently explored in strong-coupling experiments with 2D excitonic materials~\cite{Geisler:2019}.

\subsection{Cathodoluminescence spectroscopy}
The CL spectroscopy is performed in a Tescan Mira3 scanning-electron microscope (SEM) operated at an acceleration voltage of \SI{30}{\kilo\volt} and an electron beam current of \SI{2.3}{\nano\ampere}, resulting in a spot size of \SI{7.1}{\nano\meter}. Light emitted from the sample is collected by a parabolic mirror and analyzed using a Delmic SPARC cathodoluminescence detector equipped with an Andor Newton CCD camera. All spectra are background-corrected. All CL maps were collected with activated sub-pixel scanning.

\subsection{Electron-energy loss spectroscopy}
The EEL spectroscopy (EELS) and STEM imaging are performed with an FEI Titan transmission electron microscope equipped with a monochromator and a probe corrector. The microscope is operated in monochromated STEM mode at an acceleration voltage of 120\,kV, providing a spot size of approximately 0.5\,nm and an energy resolution of 0.08\,eV (measured as the full-width at half-maximum of the zero-loss peak). The microscope is equipped with a GIF Tridiem electron energy-loss spectrometer and the Gatan DigiScan acquisition system, which records an entire EELS intensity map in approximately 20\,min, depending on the number of pixels. We use a C3 aperture size of 20\,$\mu$m, camera length of 38\,mm, entrance aperture of 2.5\,mm and spectral dispersion of 0.01\,eV per pixel in the EELS measurements. In addition, we utilize the automatic drift and dark current correction function included in the acquisition system. The individual EEL spectrum of the intensity maps (with pixel sizes typically of 1.5\,nm) are recorded with acquisition times of approximately 1\,ms.

\begin{figure}[ht!]
	\centering\includegraphics[width=\textwidth]{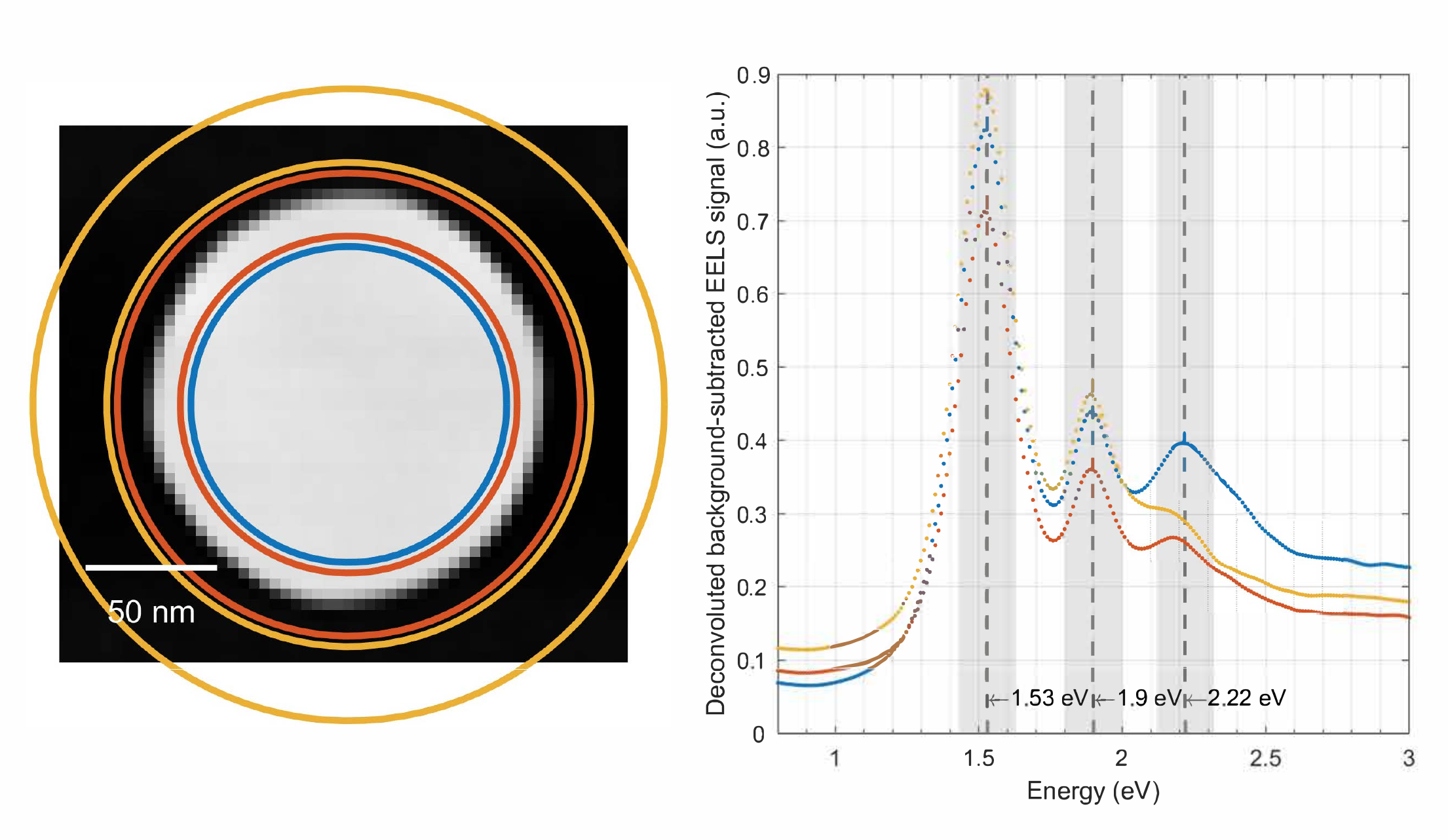}
	\caption{Analysis of EEL spectral data of a \SI{20}{\nano\meter}-thin Au nanodisk with a diameter of \SI{160}{\nano\meter}, dispersed on a \SI{20}{\nano\meter}-thin SiN substrate.}
	\label{SI:EELS}
\end{figure}

To minimize the impact of monochromator drift during acquisition, each spectrum in the EELS data matrix is normalized to the zero-loss peak. Afterwards, the zero-loss peak of each spectrum is removed using a Lucy--Richardson deconvolution algorithm, which sharpens the data by increasing the energy resolution by a factor of two approximately. We determine the number of deconvolution iterations following the procedure introduced in a previous work~\cite{Bellido:2014}. Briefly, a high-quality EEL spectrum is acquired in vacuum, which is used as the point-spread function for the deconvolution algorithm. We implement a stopping criterion, which terminates the deconvolution algorithm when the change in the reconstructed spectrum is less than 1\%. This criterion leads to between 10 to 15 deconvolution iterations. Equipped with the zero-loss profile and a fixed number of iterations, we perform the deconvolution algorithm on each spectrum in the EELS data matrix. The resonant intensity maps shown in this paper depict the summed signal in a 0.10\,eV spectral window centered at the resonance energies. The nanodisk sizes are determined using the Image Processing Toolbox in MATLAB. An example is displayed in Figure~\ref{SI:EELS}.

\subsection{Numerical calculations}

\begin{table}
	\begin{center}
  \begin{tabular}{c|ccc}
  Term 		& $\hbar\omega$ 	& $\hbar\gamma$  	& $\alpha$
  \\
  \hline
  Drude 	& 8.96\,eV	& 52\,meV	& ---
  \\
  Lorentz 1 	& 3.44\,eV	& 1.18\,eV	& 1.686
  \\
  Lorentz 2 	& 2.76\,eV	& 0.62\,eV	& 0.696
  \end{tabular}
   \caption{Parameters for our Drude--Lorentz model of gold in Equation~(\ref{eq:DL}).}
  \label{tab:Au_model}
  \end{center}
\end{table}

For comparison with theory, we performed numerical calculations of the CL signal based on the DGTD method~\cite{Busch:2011}. To this end, we used the framework that has been developed in the past for the simulation of EELS experiments~\cite{Matyssek:2011,Schroder:2015}. The system composed of the gold disk, the TEM substrate and the surrounding air was discretized in a tetrahedral mesh with element size between $80\,$nm in air away from the ND, $10\,$nm close to the ND and $5\,$nm inside the ND itself using a third-order Lagrange polynomial basis and evolved in time using a fourth--order low-storage Runge--Kutta integrator. The simulation domain had a size of $1600\,$nm and was terminated by $400\,$nm thick perfectly-matched layers. Air and silicon nitride were modelled as lossless dispersion-free dielectrics ($\varepsilon_{\text{air}} = 1.0$ and
$\varepsilon_{\text{Si}_3\text{N}_4} = 4.2025$), whereas the gold disk was modelled as a Drude--Lorentz model 
\begin{multline}\label{eq:DL}
  \varepsilon(\omega) = 
    5.8787 - \frac{\omega_p^2}{\omega (\omega + \mathrm{i}\gamma_p)} 
 \\ + \frac{\alpha_{L1} \omega_{L1}^2}{\omega_{L1}^2 - \omega^2 - \mathrm{i} \gamma_{L1}^2 \omega} 
  + \frac{\alpha_{L2} \omega_{L2}^2}{\omega_{L2}^2 - \omega^2 - \mathrm{i} \gamma_{L2}^2 \omega},
\end{multline}
that matches the value tabulated by Johnson and Christy~\cite{Johnson:1972} to within $5\%$ throughout the relevant spectral range (see Table~\ref{tab:Au_model} for parameters).

The system was excited by an electron represented by a Gaussian charge distribution ($5\,$nm wide) travelling at $0.25$ speed of light in $z$-direction using the scattered field source method~\cite{Schroder:2015}. Generated radiation (i.e. the CL signal) was recorded as the net Poynting flux through an integration contour that enclosed the metal disk and part of the substrate.

\section{Results and Discussion}

\begin{figure}[ht!]
	\centering\includegraphics[width=\textwidth]{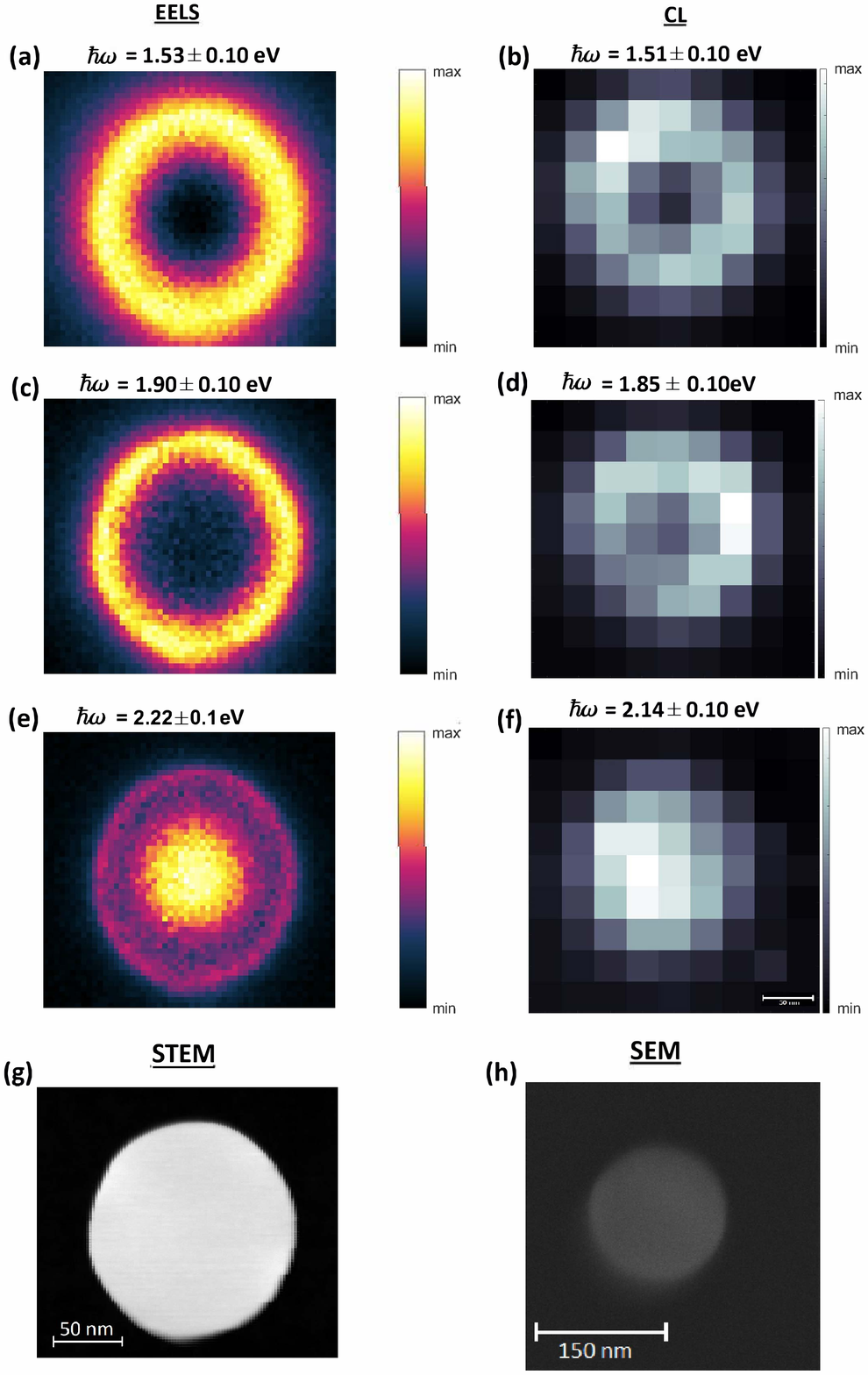}
	\caption{Spectral EELS (left) and CL (right) maps of a \SI{20}{\nano\meter} thin Au nanodisk with a diameter of \SI{160}{\nano\meter}, integrated over the spectral feature corresponding to (a) and (b) the dipolar plasmonic mode, (c) and (d) the quadrupolar plasmonic mode, and (e) and (f) the RBM. (g) and (h) show the corresponding STEM and SEM images, respectively.}
	\label{fig:EELS_CL}
\end{figure}

Figure~\ref{fig:EELS_CL} shows EEL and CL spectral maps of a \SI{20}{\nano\meter} thin monocrystalline Au ND with a diameter of $D=\SI{160}{\nano\meter}$. Each of these maps was obtained by integrating over the background-corrected peak in the EEL (cf. Figure~\ref{SI:EELS} and corresponding section) and CL spectrum, respectively. These maps reveal three plasmonic modes which can be attributed to the radiative dipolar mode at photon energies \SI[separate-uncertainty = true]{1.51(10)}{\electronvolt}, a quadrupolar mode at \SI[separate-uncertainty = true]{1.90(10)}{\electronvolt} and \SI[separate-uncertainty = true]{1.85(10)}{\electronvolt}, and a dark RBM at \SI[separate-uncertainty = true]{2.22(10)}{\electronvolt} and \SI[separate-uncertainty = true]{2.14(10)}{\electronvolt} in EELS and CL, respectively. 
We attribute the observed systematic shift between the CL and EELS resonances at least partially to the different substrate thicknesses used in both measurements.

\begin{figure}[ht!]
	\centering\includegraphics[width=\textwidth]{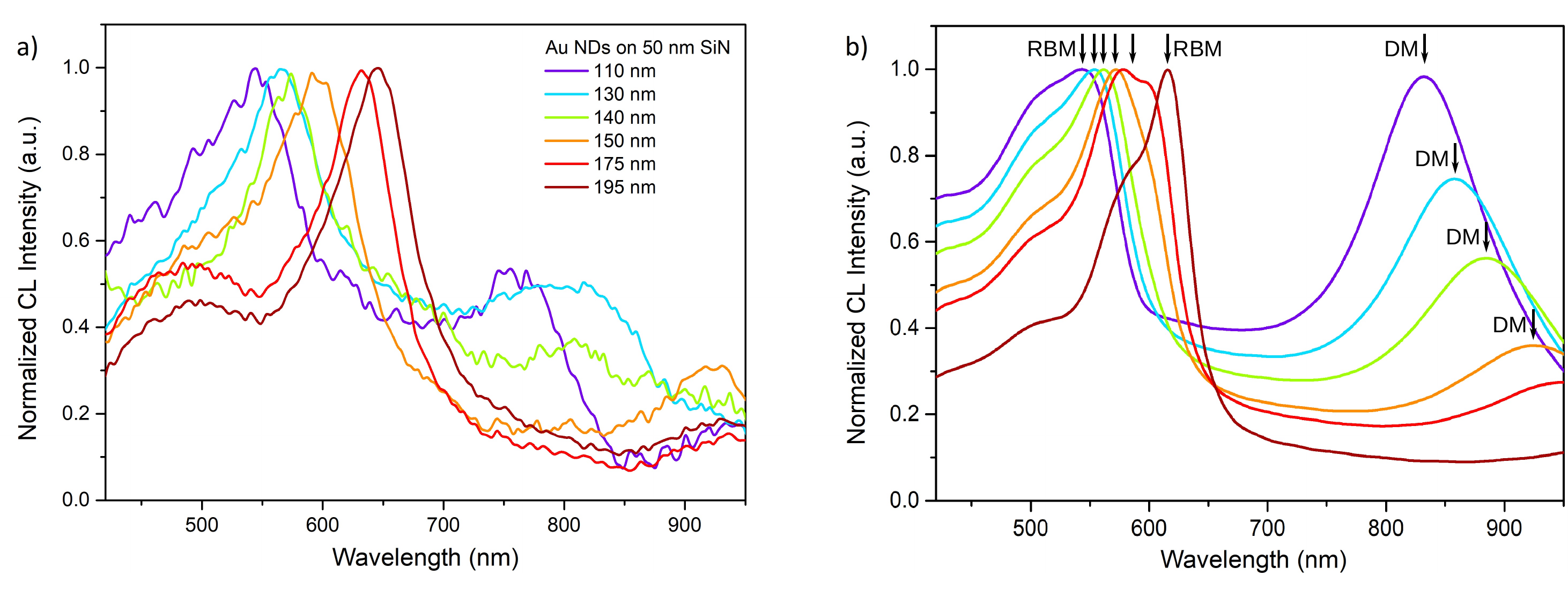}
	\caption{(a) Experimental (normalized to RBM) and (b) simulated CL spectra of \SI{20}{\nano\meter} thin Au nanodisks with varying diameter ranging from \SI{110}{\nano\meter} to \SI{195}{\nano\meter}, dispersed on a \SI{50}{\nano\meter} thin SiN membrane. Electron beam was experimentally positioned in the center of the Au ND, while it was placed \SI{5}{\nano\meter} off-center for the simulations.
	The RBMs and dipolar modes (DMs) are annotated in the numerical plots; note that the quadrupolar resonance is invisible in these plots because of the combination its low radiative visibility and the position of the electron beam next to the disk's center.}
	\label{fig:CL_dia}
\end{figure}

The most pronounced plasmonic mode supported by the Au ND is the dipole which exhibits the highest intensity in both EEL and CL spectra. The mode profile extends beyond the physical boundaries of the ND, while the less intense quadrupolar mode, radiating at a higher energy, is more confined to the rim of the ND. The dark RBM has the characteristic pattern with maximum intensity in the center and a lower intensity at the rim, separated by a dark ring. 
It is noteworthy that the difference in spatial resolution in EEL and CL spectroscopy stems from significantly longer integration times needed for the less efficient plasmonic radiation processes in CL than in EEL spectroscopy. Therefore, the pixel size in CL was reduced to \SI{25 x 25}{\nano\meter} even though the actual spot size of the electron beam was substantially smaller than that. Furthermore, the sub-pixel scanning of the electron beam was activated, therefore, an integrated CL signal over the pixel of \SI{25 x 25}{\nano\meter} is shown in the right column of Figure~\ref{fig:EELS_CL}. This not only results in a shorter collection time but also in a significantly reduced carbon deposition during the CL scan which otherwise potentially alters the measurements~\cite{Bosman:2008,Schertz:2012}. 
The three plasmonic modes identified in the EEL spectral maps are also evidently visible in the CL maps.
Note that the Au NDs in the EEL and CL measurements are not identical due to different thicknesses of the supporting SiN membranes. While in EEL a \SI{20}{\nano\meter} thin SiN membrane was used as a substrate, in CL a thicker, \SI{50}{\nano\meter}, one was chosen for better visibility which will be further discussed below.
The slight differences in shape, namely roundness, and diameter are the most likely reason why the spectral position of the three modes are close but not identical. Nevertheless, Figure~\ref{fig:EELS_CL} demonstrates that all three plasmonic modes supported by a \SI{160}{\nano\meter}-Au ND can be identified in EEL and CL, regardless of their "brightness". While the bright dipolar and quadrupolar modes are expected to be detected in CL, it is surprising that the dark RBM is also visible. As mentioned before, CL only probes the radiative part of the LDOS, therefore, modes with no effective dipole moment should, in principle, not be detectable in CL. However, this phenomenon has been reported before, using EEL and CL spectroscopy of silver NDs~\cite{Schmidt:2012, Schmidt:2018}. The authors of these studies concluded that a minimum ND diameter of \SI{200}{\nano\meter} is required to observe CL from RBMs, which was consequently attributed to retardation effects. However, our results reveal that RBMs in gold NDs of almost half the size can also generate a CL signal, indicating that the proposed explanation requires further study. We therefore systematically investigate the outcoupling properties of dark RBMs into the far-field by varying the diameter of the Au ND while keeping the rest of the parameters unchanged. 
Figure~\ref{fig:CL_dia} displays the CL spectra of Au NDs with diameters ranging from $D =$ \SI{110}{\nano\meter} to \SI{195}{\nano\meter}, dispersed  on a $t =$ \SI{50}{\nano\meter} thick SiN membrane, where the electron beam was placed in the center of the Au ND. Note that all spectra are normalized to the RBM emission.
The experimental data shown in Figure~\ref{fig:CL_dia} a) reveal a significant increase in noise level with decreasing diameter indicating that larger NDs exhibit higher CL intensity than the smaller ones. For NDs with larger diameters, the plasmonic modes are spatially more separated, therefore, the corresponding spectra are less dominated by the dipolar mode when excited by the electron beam at the center of the ND. In addition, the quasi-static surface plasmon resonance of gold, here at approximately \SI{480}{\nano\meter}, overlaps with the RBMs in smaller Au NDs which blueshift with decreasing diameter. This, and the interband transitions in gold, makes it increasingly difficult to spectrally separate the modes, which implies that the RBM can no longer be observed in Au NDs with smaller diameters than \SI{110}{\nano\meter}.

Figure~\ref{fig:CL_dia}(b) depicts the calculated CL spectra of Au NDs using the same parameters as in the experiment except for a slight offset of the simulated electron beam. As seen in Figure~\ref{tab:CL_dia}, the theoretically obtained CL spectra are in good agreement with the experiment: A similar trend of redshifting of the RBM with increasing diameter of the Au ND is visible, accompanied by an increased CL intensity. For larger Au NDs, the calculated CL spectra exhibit a second peak at approximately \SI{580}{\nano\meter} overlapping with the RBM, while no distinct features are found in the experimental data. Its origin is, at this time, not clear. Although its exploration seems to be an interesting project on its own, it is beyond the scope of this article which focuses on the outcoupling of dark plasmonic modes. 
Our results in Figure~\ref{fig:CL_dia} indicate that the spectral position of the RBM in Au NDs can be tuned over a large spectral range by varying the diameter, which makes them promising candidates for nanophotonic applications.

\begin{table}[ht!]
	\begin{center}
		\begin{tabular}{c|cc|cc} 
			Disk diameter & \multicolumn{2}{c|}{Experimental}   & \multicolumn{2}{c}{Calculated}   \\
			$D$ (\si{\nano\meter}) &  $\lambda_\textrm{exp}$ &  $\hbar\omega_\textrm{exp}$ (\si{\electronvolt}) &  $\lambda_\textrm{calc}$ (\si{\nano\meter}) &  $\hbar\omega_\textrm{calc}$ (\si{\electronvolt})\\
			\hline
			110 & 545 & 2.28 & 545 & 2.28\\
			130 & 560 & 2.21 & 555 & 2.23\\
			140 & 575 & 2.16 & 565 & 2.19\\
			150 & 595 & 2.08 & 575 & 2.16\\
			175 & 635 & 1.95 & 590 & 2.10\\
			195 & 645 & 1.92 & 615 & 2.02\\
		\end{tabular}
		\caption{Experimental and calculated emission wavelengths and photon energies of the RBMs in \SI{20}{\nano\meter} thin Au NDs with disk diameters $D$ ranging from  \SI{110}{\nano\meter} to \SI{195}{\nano\meter}.}
		\label{tab:CL_dia}
	\end{center}
\end{table}

\begin{figure}[ht!]
	\centering\includegraphics[width=\textwidth]{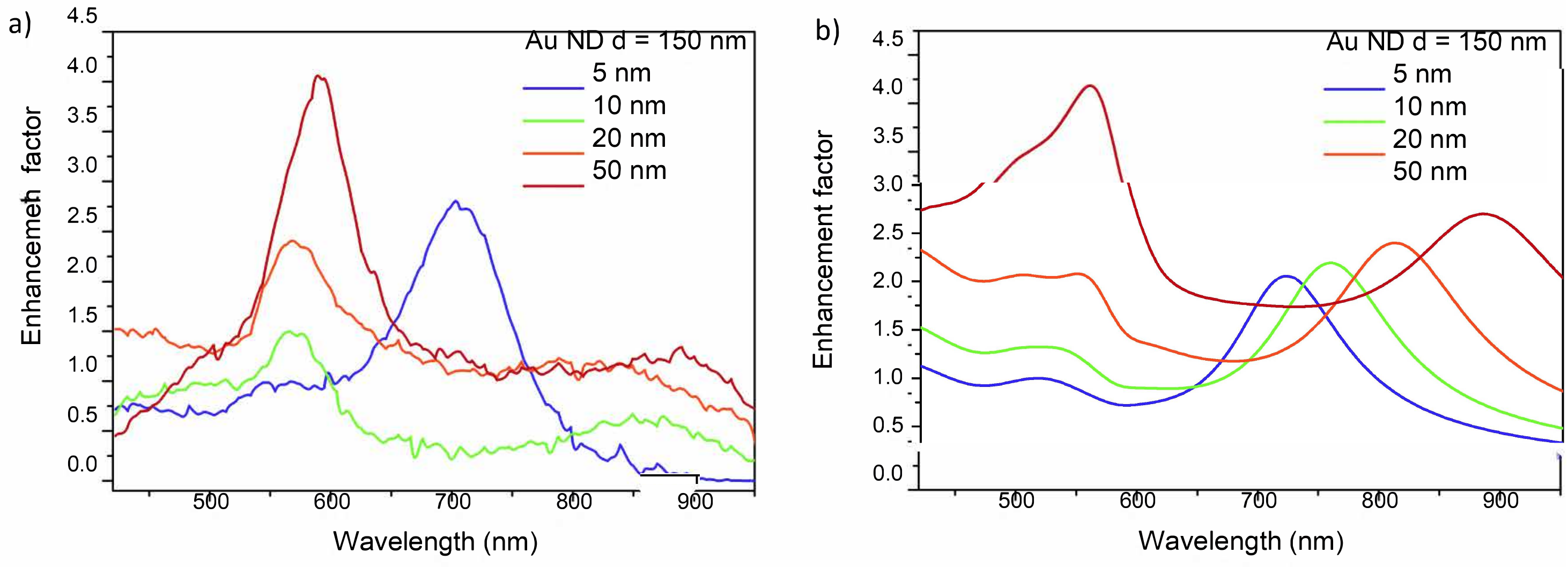}
	\caption{(a) Experimental and (b) calculated CL spectra of \SI{20}{\nano\meter} thin Au NDs with a diameter of \SI{150}{\nano\meter}, dispersed on \SI{5}{\nano\meter} to \SI{50}{\nano\meter} thin SiN substrates. Electron beam was experimentally placed in the center of the Au ND, while it was positioned \SI{10}{\nano\meter} off center for the simulations.}
	\label{fig:CL_t_SiN}
\end{figure}

To further explore the visibility of the RBM, the thickness of the SiN membrane was changed from \SI{5}{\nano\meter} to \SI{50}{\nano\meter}, while the size of the Au ND was kept constant. The resulting CL spectra are shown in Figure~\ref{fig:CL_t_SiN}. The CL intensity of the RBM gradually increases with thicker SiN membranes, which is accompanied by a spectral redshift.  

The more intense CL signal of the RBM with increasing thickness of the SiN membranes cannot be explained by a retardation effect, as this only depends on the size of the ND. However, symmetry breaking in various ways has been shown to result in radiant (or sub-radiant) modes~\cite{Wang:2006, Krug:2014, Hao:2008, Panaro:2014}, ultimately making dark modes (more) optically accessible. This concept can also be applied to the Au NDs investigated in this work:
When neglecting the substrate, they exhibit rotational symmetry along the $z$-axis, that is the direction of the electron beam, as well as mirror symmetry along the in-plane axis. Therefore, the electromagnetic field distribution of the RBM has no in-plane dipole moment, resulting in a dark, or optically inaccessible plasmonic mode. However, the high refractive index-substrate (SiN) cannot be neglected, as the electric near-field in the Au ND, induced by the electron beam, further extends into the substrate. As a result, an effective dipole moment in $z$-direction is being created, allowing radiative outcoupling of this otherwise dark mode into the far-field. The substrate, therefore, breaks the mirror symmetry of the Au ND, while its rotational symmetry is obviously preserved. 
The gradual increase of the SiN thickness gives rise to a larger amplitude of the electric field in the substrate, which leads to a larger net dipole moment, and subsequently a more intense CL signal.

\begin{figure}[ht!]
	\centering\includegraphics[width=\textwidth]{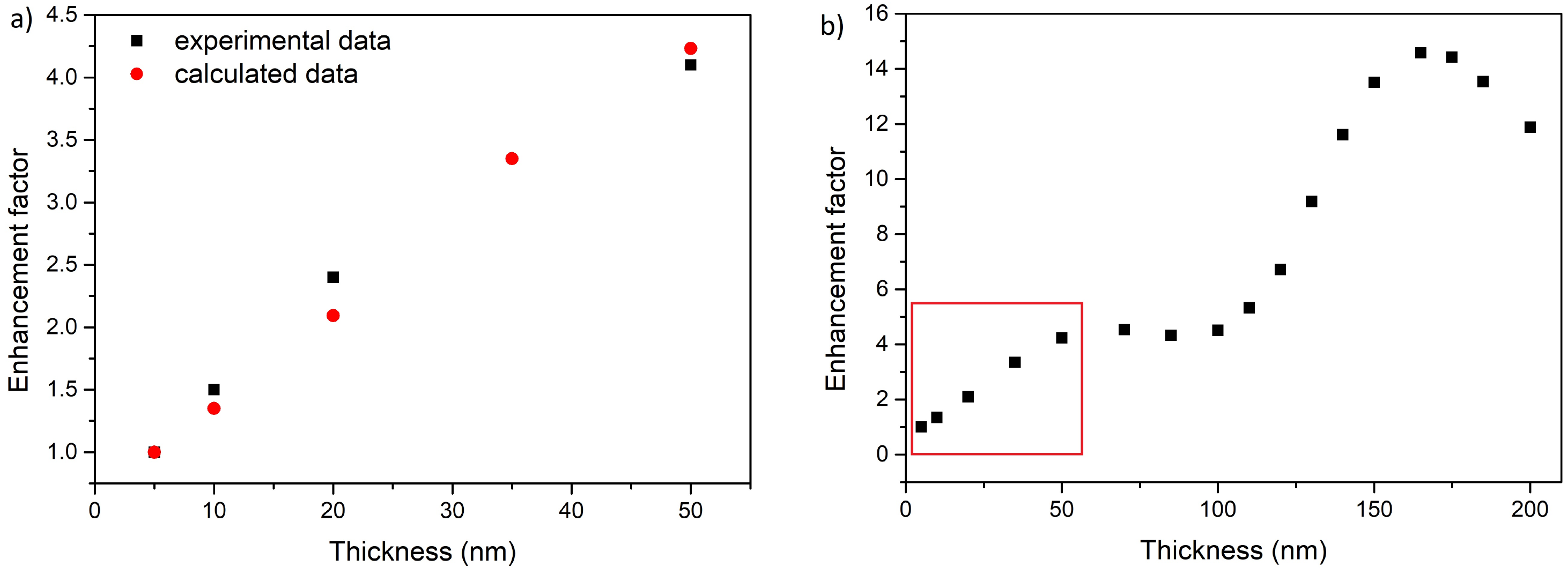}
	\caption{Maximum CL intensity of the RBM in a \SI{150}{\nano\meter}-Au ND, taken from Figure~\ref{fig:CL_t_SiN}, as a function of SiN thickness, normalized to \SI{5}{\nano\meter}-thin SiN. (a) Experimental and calculated CL data, and (b) calculated CL with extended substrate thicknesses. The peak intensities are shown relative to the value for 5\,nm substrate thickness.}
	\label{fig:CL_t_scatter}
\end{figure}

The resulting CL enhancement factor, normalized to the CL intensity of the RBM supported by a \SI{150}{\nano\meter}-Au ND on a \SI{5}{\nano\meter} thin SiN membrane, as a function of SiN thickness is shown in Figure~\ref{fig:CL_t_scatter}. An enhancement factor of more than \num{4} is found for the thickest SiN membrane of \SI{50}{\nano\meter} in the experiment which agrees very well with the corresponding calculations (cf. inset in Figure~\ref{fig:CL_t_scatter}). The simulations further reveal that the CL intensity of the RBM plateaus for thicknesses between \SI{50}{\nano\meter} and \SI{100}{\nano\meter}, before further increasing to \num{14} times for \SI{150}{\nano\meter} thick SiN, followed by a slow decrease. 
The electric near-field is considerably localized, typically at a few tens of nanometers. Therefore, the increase in CL intensity of the RBM through increasingly thicker SiN substrates is expected to be saturated at a certain thickness, which was found for \SI{50}{\nano\meter} thin SiN membranes in both, experiment and simulation.
Furthermore, we observe a second, resonant enhancement effect in our simulations for substrate thicknesses beyond 100\,nm. This is due to the hybridization of the RBM with a Fabry--P{\'e}rot resonance in the substrate. 

The resonance condition for a Purcell-enhanced RBM in a \SI{150}{\nano\meter}-Au ND, which is approximately at a wavelength of \SI{600}{\nano\meter}, is met at a SiN thickness near \SI{150}{\nano\meter}, as its refractive index is close to \num{2}.

\section{Conclusion}

In conclusion, we have studied the outcoupling behavior of radial breathing modes in small Au nanodisks into the far-field, using EEL and CL spectroscopy. These dark plasmonic modes can be excited by electron-beam irradiation and detected via a standard cathodoluminescence setup, despite collecting only the radiative decay in the far-field.
While retardation effects might play a role for larger nanodisks, the visibility of RBMs in our work can be attributed to the breaking of mirror symmetry between the top (vacuum) and bottom (high-index material substrate) of the disk. An up to \num{4}-fold CL enhancement was achieved by increasing the SiN thickness from \SI{5}{\nano\meter} to \SI{50}{\nano\meter}. Due to the limited extension of the electric near-field into the substrate, a greater enhancement is only possible by engineering Fabry--P{\'e}rot modes in the substrate to the RBM spectral position.
Our results suggest that the outcoupling of dark RBMs is not negligible. It depends strongly on the substrate thickness and therefore should be taken into account for plasmonic applications only considering the bright dipolar plasmons. 
In addition, the spectral tunability of the RBM as well as the increasing visibility achieved by choosing a suitable substrate, including its thickness, allows for their exploitation for nanophotonic applications. In particular, the possibility to take advantage of their longer lifetime, narrower linewidth and little to no radiative losses compared to their bright counterpart is an intriguing possibility for advanced realizations.

\section*{Funding}
Marie Sk\l{}odowska-Curie COFUND Action (grant agreement No. 713694); VILLUM FONDEN (grant 16498);Villum Kann Rasmussen Foundation (Award in Technical and Natural Sciences 2019); University of Southern Denmark (SDU 2020 funding); Danish National Research Foundation (CNG, project No. DNRF103). Independent Research Funding Denmark (grant no. 7026-00117B).

\section*{Acknowledgments}

We thank A.~S. Roberts and J.~Linnet for technical assistance during the installation and testing phases of the CL instrumentation. We also thank S.~Boroviks and C.~Tserkezis for stimulating discussions.

\bibliography{sample}

\onecolumngrid

\end{document}